\documentclass[a4paper,11pt]{article}
\usepackage{pos}
\newcommand{\vphi}{\vec{\phi}}
\newcommand{\vphisq}{\vec{\phi}^{\, 2}}
\newcommand{\tilmom}{\widetilde{m}_\omega}

\title{How an $\omega_0$ condensate can spike the speed of sound in cold quarkyonic matter}
\ShortTitle{Spiking the speed of sound with an $\omega_0$ condensate}

\author*[a]{Robert D. Pisarski}
\affiliation[a]{Dept. of Physics, Brookhaven National Laboratory,\\
  Upton, NY, 11973, USA}

\emailAdd{pisarski@bnl.gov}

\abstract{I describe a novel mechanism where the variation of an $\omega_0$ condensate can generate
  a ``spike'' in the speed of sound in hadronic matter.  An $\omega_0$ condensate naturally increases
  the speed of sound; the real problem is how to get the speed of sound to decrease.  I suggest this can happen
  through the appearance of a Quantum Pion Liquid.}

\FullConference{%
	The International conference on Critical Point and Onset of Deconfinement - CPOD2021\\
  	15 – 19 March 2021\\
 	Online - zoom
 }

\begin{document}
\maketitle

Consider a $SU(N_c)$ non-Abelian gauge theory, coupled to $N_f$ flavors of quarks in the fundamental representation,
at nonzero temperature and zero quark chemical potential.  As the temperature is raised, theoretically
there are two phase transitions possible.  One is deconfinement, and for light quarks, a second for the
restoration of chiral symmetry.

The details of these transitions depend upon the number colors and flavors.  One particularly simple example is the case
of an infinite number of colors, where the number of flavors is held fixed as $N_c \rightarrow \infty$.  Then everything
is dominated by the pure glue theory, and we expect (and numerical simulations on the lattice confirm) that the
deconfining transition, at a temperature $T_d$, is of first order.
This is easy to understand.  In the confined phase all hadrons are color singlets, and
so any hadron has a degeneracy of order one.  Thus the pressure in the confined phase, below $T_d$, is
of order one.  In the deconfined phase, above $T_d$ we can certainly have a complicated, strongly interacting
phase until very high temperature.  Even so, no matter how strongly the deconfined quarks and gluons interact,
there are $\sim N_c N_f$ quarks and $\sim N_c^2$ gluons, with a pressure which is $\sim N_c^2$.  That is,
in the limit of infinite $N_c$ the pressure itself can be used as an order parameter.

If there are also massless quarks, we can also characterize the restoration of chiral symmetry at a temperature
$T_\chi$.  If the number of flavors is held finite as $N_c \rightarrow \infty$, though, the quarks are really
driven by the dynamics in the pure glue theory.  Thus it is very hard to imagine that $T_\chi < T_d$: why
should the quark dynamics change at all?

Indeed, general arguments suggest that the confined phase is completely
independent of temperature.  At first sight this sounds surprising, but again it isn't.  Not only are
the number of hadrons of order one, but any interactions between them
are suppressed by powers of $1/N_c$.  Thus in the strict limit of
infinite $N_c$, there is no way that any quantity can change with temperature.  This
is why a first order transition is expected: the confined phase has a pressure of one, and the deconfined
phase has a pressure which is negative below $T_d$ and positive above.  Thus the
derivative of the pressure is nonzero at $T_d$, which implies that the energy density jumps from
$\sim 1$ below $T_d$ to $\sim N_c^2$ above.  If one works hard, one can use the Hagedorn spectrum
to get a second order deconfining transition at infinite $N_c$ \cite{Pisarski:1983db}, but as I said, the numerical
evidence strongly disfavors this.

Similarly, it is possible that $T_\chi$ is greater than $T_d$, but again, it would
be unexpected.  Instead, the safest best is simply that when deconfinement occurs, chiral restoration also occurs,
$T_\chi = T_d$.

Let me make another, apparently trivial, comment.  At zero quark chemical potential, in the confined
phase there is no condensate which affects deconfinement.  The order parameter for deconfinement
is zero below $T_d$, and is only nonzero above $T_d$.  Of course below $T_d$ the chiral condensate,
$\langle \overline{\psi} \psi \rangle$, is nonzero.  Still, any loop which contributes to a change in the chiral
condensate with temperature involves a coupling to the hadronic state, and so the change will be suppressed by
some power of $1/N_c$.

All of these clean results are special to holding $N_f$ fixed as $N_c \rightarrow \infty$.  If $N_f$ is as large
as $N_c$, then most of the above results go away.  In particular, since there are $\sim N_f^2$ hadronic states,
then we don't even know the order of the transition: it could be first order, but it could be second, or crossover.
The pressure is then $\sim N_f^2 \sim N_c^2$ at all temperatures.  Similarly, if neither $N_f$ nor $N_c$ is large,
then we need the results of numerical simulations on the lattice.

Let us then return to the case of $N_f \ll N_c \rightarrow \infty$, and consider nonzero quark chemical potential,
$\mu_{qk}$.  For the pressure, the quark contribution is no larger than $\sim N_f N_c \mu^4_{qk}$, so
at nonzero temperature, the quark contribution is only commesurate with that from gluons, $\sim N_c^2 T^4$, when 
$\mu_{qk} \sim N_c^{1/4} T$.    This is the basic idea behind a quarkyonic regime at nonzero density,
where the free energy is that of (interacting) quarks and gluons, but excitations near the Fermi surface
are still confined
\cite{McLerran:2007qj,Kojo:2009ha,Kojo:2010fe,Kojo:2011cn,McLerran:2018hbz,Pisarski:2021aoz,Pisarski:2021qof,Tsvelik:2021ccp}.

However, unlike the case of nonzero temperature, the confined phase can depend upon density.  This is because
while couplings between mesons are suppressed at large $N_c$, those between mesons and baryons are not: they
are large, $\sim N_c^{1/2}$.  This suggests that it is possible for the chiral transition to {\it split} from
the deconfining transition at nonzero $\mu_{qk}$, so that as $\mu_{qk}$ increases, $T_\chi$ is {\it less} than
the deconfining temperature.  Of course $\mu_{qk}$ only matters when one can first generate a Fermi sea of
baryons, which requires $\mu_{qk} > m_B/N_c$, where $m_B$ is the mass of the lightest baryon.

Further, at nonzero density there is uniquely one condensate which plays a privileged role.  The importance of
this was first noted by Zeldovich \cite{Zeldovich:1962}; it also provides
the basis for the effective theory of nuclear matter, Quantum
HadroDynamics \cite{Serot:1997xg}.  At nonzero density, by definition there is a nonzero
expectation value for the timelike component of the current for fermion number,
$\langle \overline{\psi} \gamma^0 \psi \rangle \neq 0$.  But this current couples directly to that for
the $\omega_0$ meson, as that couples to nucleons as
$g_\omega \overline{\psi}^\dagger \gamma^\mu \omega_\mu \psi $.  Thus if $n_B$ is the baryon density,
nonzero density generates a term linear in $\omega_0$, and induces an expectation value for $\omega_0$:
\begin{equation}
  {\cal L}_\omega^B = - g_\omega n_B \omega_0 + \frac{m_\omega^2 \omega_\mu^2}{2} \Rightarrow
  \langle \omega_0 \rangle = \frac{g_\omega }{m_\omega^2} \, n_B \; .
  \label{omega_lag}
\end{equation}
Here $m_\omega$ is the mass of the $\omega$ meson.  With $n_B \sim N_c^0$ and $g_\omega N_c^{1/2}$, this
contributes to the free energy as $\sim N_c$, as expected for the quark contribution.

The mass of the $\omega$ meson can be nontrivial.  I take the effective Lagrangian for the $\omega$ meson as
\begin{equation}
  {\cal L}_\omega = \frac{{\cal F}_{\mu \nu}^2 }{4} +
  \frac{1}{2} \left( \tilmom^2 + \kappa^2 \vphisq \right) \omega_\mu^2   \; .
  \label{omega_lag_mass}
\end{equation}
${\cal F}_{\mu \nu} = \partial_\mu \omega_\nu - \partial_\nu \omega_\mu$ is the
standard, Abelian field strength for the $\omega_\mu$ meson.  The mass
$\tilmom$ is a constant, but in addition I also add a quartic coupling, $\sim \kappa^2$, between
$\omega_\mu$.  Here
$\vphi$ is the $O(4)$ chiral field for two light flavors, $\vphi = (\sigma,\vec{\pi})$.
The coupling $\kappa^2$ must be positive to ensure stability for large values of the $\omega_\mu$ and $\vphi$ fields.

In the vacuum, the $\sigma$ field acquires a vacuum expectation value from chiral symmetry breaking,
$\langle \sigma \rangle = f_\pi$, so that the $\omega_\mu$ mass is $m_\omega^2 = \tilmom^2 + \kappa f_\pi^2$.
In vacuum this is fine, as one can't really distinguish between the part of the $\omega_\mu$ mass which is
bare, and the part induced by chiral symmetry breaking.  Similarly, vector meson dominance is used
to characterize the coupling of the $\rho_\mu$ meson to the photon, but it doesn't constrain how the
$\rho_\mu$ mass arises.

At the stationary point in $\omega_0$, the effective Lagrangian of Eq. (\ref{omega_lag}) becomes
\begin{equation}
  {\cal L}_\omega^B = - \frac{g_\omega}{m_\omega^2}\;  n_B^2 \; .
  \label{omega_lag2}
\end{equation}
As demonstrated first by Zeldovich \cite{Zeldovich:1962}, this gives a speed of sound equal to the speed of light.  This
also arises at nonzero isospin density, in the limit of asymptotically large density.

Of course this is only a leading term, and cannot be taken as exact.  Instead, one should use a model of
Quantum HadroDynamics (QHD) \cite{Serot:1997xg}, where saturation arises
from a balance between attraction, due to exchange of a $\sigma$-meson, and repulsion, from exchange
of the $\omega_\mu$ meson.  This balance tends to weaken the effect of the leading order term in
Eq. (\ref{omega_lag2}), but still, the speed of sound tends to increase strongly.

This is important for astrophysics.
Drischler {\it et al.} \cite{Drischler:2020fvz} use chiral effective field theory
to extrapolate from the saturation density of nuclear matter, $n_0$, to twice that.
Given the observation of neutron stars with masses above two solar masses
\cite{Demorest:2010bx,Antoniadis:2013pzd}, it is {\it imperative} to have a region of density
in which the EoS is stiff, with a speed of sound significantly above that
of an ideal quark gas, where $c_s^2 = 1/3$.  
Using the small tidal deformability observed from GW170817, though, the EoS of nuclear matter must be soft until
$n_B \sim 1.5 - 1.8 n_0$, and then increase sharply.
That is, there is a ``spike'' in the speed of sound, with a relatively narrow peak at a density 
significantly above $n_0$, Fig. (1) of Ref. \cite{Greif:2018njt}.

As I argued above, an $\omega_0$ condensate naturally gives an {\it in}crease in the speed of sound.  The question
which I wish to stress is the following: {\it how can one get the $\omega_0$ condensate to evaporate}, and thus
for the speed of sound to {\it de}crease?

In a quarkyonic phase, which is confined, it is manifestly sensible to speak of an $\omega_\mu$ meson.
This is very different from the case of increasing temperature at zero density: then there is no condensate
for the $\omega_0$, and vector mesons just fall apart into the constituent quarks.  But at nonzero density,
in the quarkyonic phase the $\omega_\mu$ meson remains confined, and as a confined meson, can't fall apart into
quarks.

This problem does not appear to have been appreciated previously.  Either a QHD-type result was used uniformly,
or a QHD model was matched onto constituent quarks, as in the model of Cao and Liao \cite{Cao:2020byn}.  But
the $\omega_\mu$ meson doesn't go away: there must be a {\it dynamical} reason why the contribution of the
$\omega_0$ condensate evaporates at nonzero density.

I suggest a rather speculative argument as to how this could happen.  Consider the effective mass of the $\omega_\mu$
meson, including the $\kappa$ coupling:
\begin{equation}
 \langle \omega_0 \rangle  = \frac{g_\omega \rho_B}{\tilmom^2 + \kappa^2 \langle \phi^2 \rangle} \; .
\end{equation}
As the chiral transition is approached, this will tend to increase, as the chiral condensate decreases,
$\langle \phi^2 \rangle \approx \langle \sigma \rangle^2$.  This stiffens the equation of state.

The simplest way for the $\omega_0$ condensate to become small is for the $\omega_\mu$ meson to become
large.  This is where the $\kappa$ coupling enters.  In a linear $\sigma$ model \cite{Pisarski:2020dnx},
this can arise as follows.  Consider an effective chiral Lagrangian,
\begin{equation}
  {\cal L}_\phi = \frac{1}{2} (\partial_0 \vphi )^2 + \frac{1}{2 M^2} ( \partial_i^2 \vphi)^2
  + \frac{Z}{2} ( \partial_i \vphi )^2
  + \frac{m_0^2}{2} \vphisq
  + \frac{\lambda }{4} (\vphisq)^2 \; .
  \label{lag_phi}
\end{equation}
This includes higher spatial derivatives, but by causality, only two time derivatives.  If
the coefficient of the term with two spatial derivatives, $Z$, is negative, in mean field
theory spatially inhomogeneous structures (``chiral spirals'') are generated.
The simplest chiral spiral is one where although $\vphi$ winds along a single spatial direction,
that $\vphisq$ is constant (a ``single mode'').  This is no problem, {\it unless} 
there are Goldstone bosons.  In that case, the Goldstone bosons have zero energy
at a {\it non}-zero spatial momentum, $k_0$:
\begin{equation}
  \Delta^{-1}_{\rm transverse}(\omega,\vec{k})
  = \omega^2 + \frac{1}{M^2} (\vec{k}^2)^2 + Z \vec{k}^2 + m^2 + \lambda \phi_c^2
  = \omega^2 + \frac{(k - k_0)^2}{M^2} \; , 
\end{equation}
In isospin space, the Goldstone bosons are transverse in isospin space to the
field for the background chiral spiral, $\vphi_c$,
whose period is $2 \pi/|k_0|$, where $k_0^2 \sim - Z M^2$.  However,
consider the tadpole diagram for fluctuations over the Goldstone modes.  For simplicity,
I consider the integral at nonzero temperature, taking the static mode in energy, and the
integral over the three spatial momenta \cite{Pisarski:2020dnx}:
\begin{equation}
  \delta m^2 \sim \lambda \langle \vphisq \rangle \sim \lambda \, T \int d^3 k \;  \frac{1}{(k - k_0)^2}
  \sim \lambda T k_0^2 \int \frac{ d \delta k}{(\delta k)^2} \; .
\end{equation}  
Because the Goldstone boson has zero energy at nonzero momentum, there is a {\it linear}
infrared divergence, about the momentum for the chiral spiral.  (This happens in any number of
spatial dimensions.)
The only way to avoid this divergence is if a {\it dynamical} mass is generated for
the Goldstone bosons, which adds a term $m_{\rm dyn}^2$ to the propagator.
This is a non-perturbative phenomenon, and generates a novel form of the symmetric phase, 
which in Ref. \cite{Pisarski:2021aoz} I termed a ``quantum pion liquid'' (Q$\pi$L).
(In Ref. \cite{Pisarski:2020dnx} we used the term a pion quantum spin liquid, which isn't as accurate,
since pions don't carry spin.)
In this Q$\pi$L, one can show that the
fluctuations are {\it very} large at large $Z$, $\langle \vphisq \rangle \sim Z^2$.  This means that the $\omega_\mu$
mass squared is $m_\omega^2 \sim Z^2$, so $\langle \omega_0 \rangle \sim 1/Z^2$.

At zero temperature, after integrating over $\omega$ the integral over spatial momenta gives
a logarithmic instead of a linear infrared divergence
\cite{Pisarski:2021aoz}, so that the dynamical mass is exponentially small.  
The magnitude of the fluctuations remains the same,
$\langle \vphisq \rangle \sim Z^2$, because a much smaller dynamical
mass compensates the weaker infrared divergence.
As a non-perturbative result, to demonstrate this we considered
an $O(N)$ model, and solving in the limit of infinite $N$.
It is not clear if it is valid for $N=4$, but it is reasonable to suspect that it is.

This leaves a raft of questions open.  The analysis of the sigma model assumes that when $Z < 0$, that
$\langle \vphisq \rangle$ is constant.  Under this assumption, it is easy to show that the energy
of the Goldstone bosons vanishes at a nonzero momentum, characteristic of the would be momentum of the
spatially inhomogeneous condensate.  This appears to be a technical assumption, but certainly one
expects that the energy of the Goldstone bosons will vanish at {\it some} momentum.  If $Z$ is positive,
then it is certainly at zero momentum.  When $Z$ is negative, however, it is natural to expect that
the zero of the energy is at nonzero momentum, which naturally generates a Q$\pi$L.  

There are many other avenues of investigation.
Couplings similar to $\kappa$, which couple chiral and vector mesons, have been introduced by
Refs. \cite{Dexheimer:2018dhb}.  It is imperative to consider these
couplings, as a neutron star is not isospin symmetric.

In the end, the most direct conclusion is the following.  The standard assumption is that
in the plane of temperature and chemical potential, that there is a single transition which is something
like a semi-circle.  Instead, the example of small $N_f$ and large $N_c$ suggests that cold, dense
quark matter may look {\it nothing} like the deconfined phase at zero density.  Understanding the
properties of such nuclear matter brings together results from condensed matter to astrophysics.

%\bibliographystyle{JHEP}
%\bibliography{lifshitz.bib}

\providecommand{\href}[2]{#2}\begingroup\raggedright\endgroup

%\begin{thebibliography}{99}
%\bibitem{...}
%\end{thebibliography}

\end{document}